\documentclass{elsart}
\pdfoutput=1

\usepackage{natbib}

\usepackage{graphicx}

\usepackage{amssymb}
\usepackage{amsmath}

 \usepackage{lineno}

\linenumbers
\begin{document}

\begin{frontmatter}



\title{Estimating black carbon aging time-scales with a particle-resolved
  aerosol model}


\author[label1]{Nicole Riemer\corauthref{cor1}\ead{nriemer@illinois.edu}}, \author[label2]{Matthew West},
\author[label3]{Rahul Zaveri}, \author[label3]{Richard Easter}

\address[label1]{Department of Atmospheric Science, University of
  Illinois at Urbana-Champaign, Urbana, Illinois, USA}
\address[label2]{Department of Mechanical Science and Engineering,
  University of Illinois at Urbana-Champaign, Urbana, Illinois, USA}
\address[label3]{Atmospheric Science and Global Change Division,
  Pacific Northwest National Laboratory, Richland, Washington, USA}
\corauth[cor1]{Corresponding author}

\begin{abstract}
  Understanding the aging process of aerosol particles is important
  for assessing their chemical reactivity, cloud condensation nuclei
  activity, radiative properties and health impacts. In this study we
  investigate the aging of black carbon containing particles in an
  idealized urban plume using a new approach, the particle-resolved
  aerosol model PartMC-MOSAIC. We present a method to estimate aging
  time-scales using an aging criterion based on cloud condensation
  nuclei activation. The results show a separation into a daytime
  regime where condensation dominates and a nighttime regime where
  coagulation dominates. For the chosen urban plume scenario,
  depending on the supersaturation threshold, the values for the aging
  time-scales vary between 0.06~hours and 10~hours during the day, and
  between 6~hours and 20~hours during the night.
\end{abstract}

\begin{keyword}
black carbon \sep aerosol aging \sep mixing state \sep CCN

\end{keyword}

\end{frontmatter}

\section{Introduction}
\label{sec:introduction}

Black carbon containing particles, or ``soot'' particles, are
ubiquitous in the atmosphere and their role for regional and global
climate has been widely recognized \citep{IPCC2007}. Since black
carbon absorbs light \citep{Horvath1993}, it contributes to the
aerosol radiative forcing, potentially partially offsetting the
cooling effect of scattering aerosol particles such as sulfates
\citep{Menon2002}.

Black carbon containing particles originate from the incomplete
combustion of carbon containing material, hence emissions from traffic
are an important contributor \citep{Bond2004}. Other important sources
for black carbon include biomass burning and the combustion of coal by
industrial processes. In this paper we focus on black carbon from
traffic emissions. Measurements of vehicle emissions from gasoline and
diesel cars show that the emitted particles are a complex mixture of
many chemical species with the main constituents being black carbon
and organic carbon \citep{Medalia1982, Toner2006}. Trace
concentrations of ionic and metallic species are also present
\citep{Kleeman2000}. The exact composition depends on several factors,
including the fuel type, the operating conditions and the condition of
the individual vehicles.

During their transport in the atmosphere, the composition of these
particle emissions are further modified. Coagulation, condensation and
photochemistry are contributing processes, collectively known as aging
\citep{Weingartner1997}.  During this aging process the composition of
the individual particles or, in other words, their mixing states
change \citep{Furutani2008}. This impacts the particles'
physico-chemical properties including their chemical reactivity,
radiative properties and health impacts. In particular, the aging
process can change the particles' hygroscopicity from initially
hydrophobic to more hydrophylic, and hence change their ability to
become cloud condensation nuclei
\citep{McFiggans2006,McMurry1989,Moffet2008,Cubison2008}.

This is important as models and observations suggest that wet
deposition represents 70--85\% of the tropospheric sink for
carbonaceous aerosol mass \citep{Poeschl2005}. As a consequence, to
assess the budget and impact of black carbon, models need to capture
the aging process adequately. Many global models have simulated both
(fresh) hydrophobic black carbon and (aged) hydrophilic black carbon,
which can be considered a minimal representation of the black carbon
mixing state \citep{Cooke1999, Lohmann1999, Koch2001, Croft2005}. In
such a framework only the hydrophilic black carbon is subject to
in-cloud scavenging. The conversion from hydrophobic to hydrophilic is
frequently modeled as a first-order system with the single parameter
of aging rate or its inverse, the aging time-scale $\tau$ which
represents the time-scale on which a population of black carbon
containing particles transfers from the ``fresh'' category to the
``aged'' category.

While conceptually simple, the actual value of the aging time-scale
$\tau$ is not well constrained. \citet{Koch2001} and \citet{Croft2005}
compared different aging parameterizations in global models and
concluded that the model results critically depended on the respective
formulation.  \citet{Riemer2004} used mesoscale simulations to
determine $\tau$. They derived the aging time-scale for black carbon
particles as a function of height and time of the day, which suggested
that assuming a single parameter for the black carbon aging time-scale
is an oversimplification that will incorrectly estimate the black
carbon burden. However, even though this treatment allowed more
detailed insight into the aging process, it was still based on ad hoc
aging rules inherent to the modal model framework that was used
\citep{Riemer2003}.

Recently, \citet{Riemer2009} developed a particle-resolved aerosol
model, PartMC-MOSAIC, which explicitly resolves the composition of
individual particles in a given population of different types of
aerosol particles, so that no ad hoc aging criteria needs to be
invoked. They applied PartMC-MOSAIC in a Lagrangian box-model
framework to an idealized urban plume scenario to study the evolution
of urban aerosols due to coagulation and condensation over the course
of 24~hours.

In this study, we build upon \citet{Riemer2009} and present a method
for estimating aging time-scales of black carbon containing particles
using PartMC-MOSAIC, based on the idealized urban plume scenario. We
take the particle population simulated in \citet{Riemer2009} and use a
CCN-based aging criteria to determine whether each individual particle
is fresh or aged at every timestep, and how many particles transfer
between the fresh and aged categories during each timestep. By fitting
these results to a first-order bulk model of aerosol aging we are able
to determine the aging timescale without making any a priori
assumptions about the aging process. To our knowledge it is the first
time that a method is presented for explicitly calculating aging
time-scales.

Section~\ref{sec:model} introduces the model system. In
Section~\ref{sec:urban_plume} we describe the idealized plume scenario
that served as a basis for the time-scale
estimation. Section~\ref{sec:aging_model} presents our method for
deriving time-scales from our model and Section~\ref{sec:results}
shows the results. We summarize our findings in
Section~\ref{sec:conclusion}.

\section{Model description}
\label{sec:model}

PartMC-MOSAIC is a particle-resolved model that simulates the
evolution of individual aerosol particles and trace gases in a single
parcel (or volume) of air moving along a specified trajectory. For
each particle the mass of each constituent species is tracked, but the
particle position in space is not simulated, making this a
zero-dimensional or box model. In addition to coagulation and aerosol-
and gas-phase chemistry, the model includes prescribed emissions of
aerosols and gases, and mixing of the parcel with background air. The
simulation results shown here use around 100,000 particles in a volume
of around $16\rm\,cm^{3}$ (the precise values vary over the course of
the simulation). We regard this volume as being representative of a
much larger air parcel. The model accurately predicts both number and
mass size distributions and is therefore suited for applications where
either quantity is required. Details of PartMC-MOSAIC and the urban
plume scenario are described in \citet{Riemer2009}. Here we give a
brief summary.

The simulation of the aerosol state proceeds by two mechanisms. First,
the composition of each particle can change as species condense from
the gas phase and evaporate to it. Second, the aerosol population can
have particles added and removed, either by coagulation events between
particles, by emissions, or by dilution. While
condensation/evaporation is handled deterministically, emission,
dilution and coagulation are treated with a stochastic approach.

Coagulation between aerosol particles is simulated in PartMC by
generating a realization of a Poisson process with a Brownian
coagulation kernel. For the large number of particles used here it is
necessary to employ an efficient approximate simulation method. We
developed a binned sampling method to efficiently sample from the
highly multi-scale coagulation kernel (in our case the Brownian
kernel) in the presence of a very non-uniform particle size
distribution, which is described in detail in \citet{Riemer2009}.

Particle emissions and dilution with background air are also
implemented in a stochastic manner. Because we are using a finite
number of particles to approximate the current aerosol population, we
need to add a finite number of emitted particles to the volume at each
timestep. Over time these finite particle samplings should approximate
the continuum emission distribution, so the samplings at each timestep
must be different. As for coagulation, we assume that emissions are
memoryless, so that emission of each particle is uncorrelated with
emission of any other particle. Under this assumption the appropriate
statistics are Poisson distributed, whereby the distribution of finite
particles is parametrized by the mean emission rate and distribution.

Lastly, we must also obtain a finite sampling of background particles
that have diluted into our computational volume during each
timestep. In addition, some of the particles in our current sample
will dilute out of our volume and will be lost, so this must be
sampled as well. Again, we assume that dilution is memoryless, so that
dilution of each particle is uncorrelated with the dilution of any
other particle or itself at other times, and that once a particle
dilutes out it is lost.

We coupled the stochastic PartMC particle-resolved aerosol model to
the deterministic MOSAIC gas- and aerosol-chemistry code
\citep{Zaveri2008} in a time- or operator-splitting fashion
\citep[Section~20.3.3]{NR2007}. MOSAIC treats all the globally
important aerosol species including sulfate, nitrate, chloride,
carbonate, ammonium, sodium, calcium, primary organic aerosol (POA),
secondary organic aerosol (SOA), black carbon (BC), and inert
inorganic mass.

MOSAIC consists of four computationally efficient modules: 1) the
gas-phase photochemical mechanism CBM-Z \citep{Zaveri1999}; 2) the
Multicomponent Taylor Expansion Method (MTEM) for estimating activity
coefficients of electrolytes and ions in aqueous solutions
\citep{Zaveri2005a}; 3) the Multicomponent Equilibrium Solver for
Aerosols (MESA) for intra-particle solid-liquid partitioning
\citep{Zaveri2005b}; and 4) the Adaptive Step Time-split Euler Method
(ASTEM) for dynamic gas-particle partitioning over size- and
composition-resolved aerosol \citep{Zaveri2008}. The version of MOSAIC
implemented here also includes a treatment for SOA based on the SORGAM
scheme \citep{Schell2001}.

\section{Idealized urban plume scenario}
\label{sec:urban_plume}

\begin{figure}
  \begin{center}
    \includegraphics{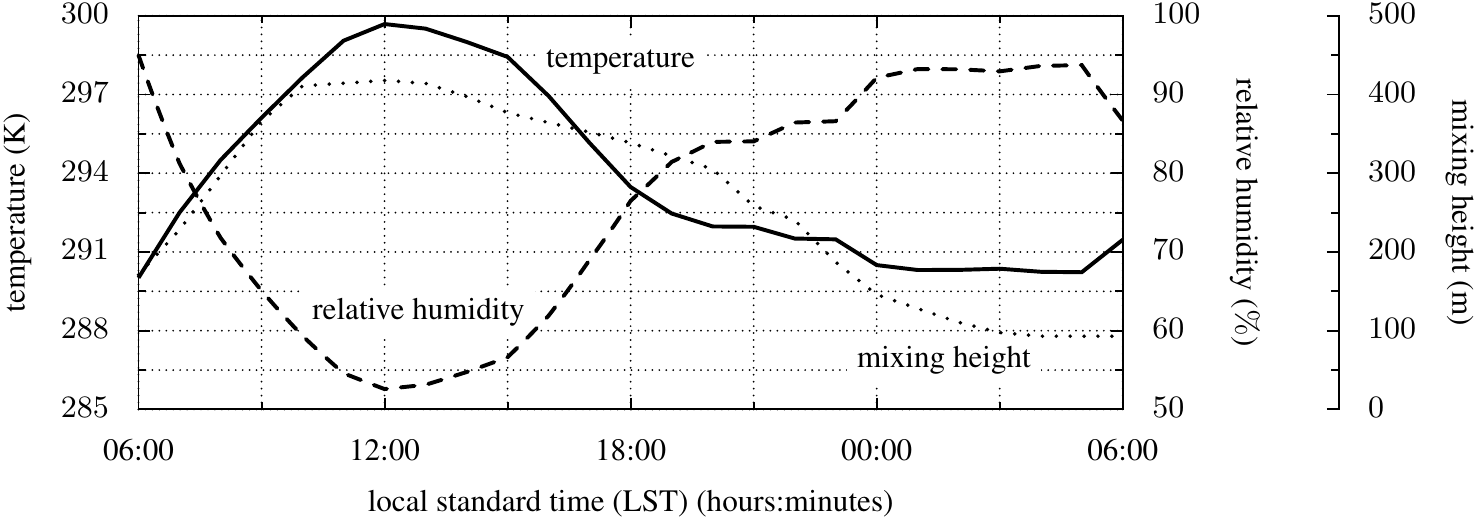}
  \end{center}
  \caption{\label{fig:aging_env} Time series of temperature, relative
    humidity, and mixing height over the course of the 24~hour
    simulation. The pressure and water mixing ratio were kept
    constant.}
\end{figure}

For our urban plume scenario, we tracked the evolution of gas phase
species and aerosol particles in a Lagrangian air parcel that
initially contained background air and was advected over and beyond a
large urban area, as described in \citet{Riemer2009}. The simulation
started at 06:00 local standard time (LST), and during the advection
process, primary trace gases and aerosol particles from different
sources were emitted into the air parcel for the duration of
12~hours. After 18:00~LST, all emissions were switched off, and the
evolution of the air parcel was tracked for another 12~hours.  The
time series of temperature, relative humidity and mixing height are
shown in Figure~\ref{fig:aging_env}.

Initial gas-phase and aerosol particle concentrations as well as gas
phase and particle emissions were the same as in
\citet{Riemer2009}. The gas phase emissions varied throughout the
emission time interval according to a typical diurnal cycle found in
polluted urban areas.

The initial particle distribution, which was identical to the
background aerosol distribution, was bimodal with Aitken and
accumulation modes \citep{Jaenicke1993}. We assumed that it consisted
of $\rm (NH_4)_2SO_4$ and POA, as shown in
Table~\ref{tab:aero_dat}. We considered three different types of
carbonaceous aerosol emissions: 1)~meat cooking aerosol, 2)~diesel
vehicle emissions, and 3)~gasoline vehicle emissions. The parameters
for the distributions of these three emission categories were based on
\citet{Eldering1996}, \citet{Kittelson2006-1}, and
\citet{Kittelson2006-2}, respectively. For simplicity in this
idealized study, the particle emissions strength and their size
distribution and composition were kept constant with time during the
time period of emission.

Furthermore, we assumed that every particle from a given source had
the same composition, with the species listed in
Table~\ref{tab:aero_dat}, since to date the mixing state of particle
emissions is still not well quantified. In particular, we assume that
the diesel and gasoline exhaust particles consist exclusively of POA
and BC, which is very nearly the case
\citep{Andreae2006,Medalia1982,Kleeman2000}.

\begin{figure}
  \begin{center}
    \includegraphics{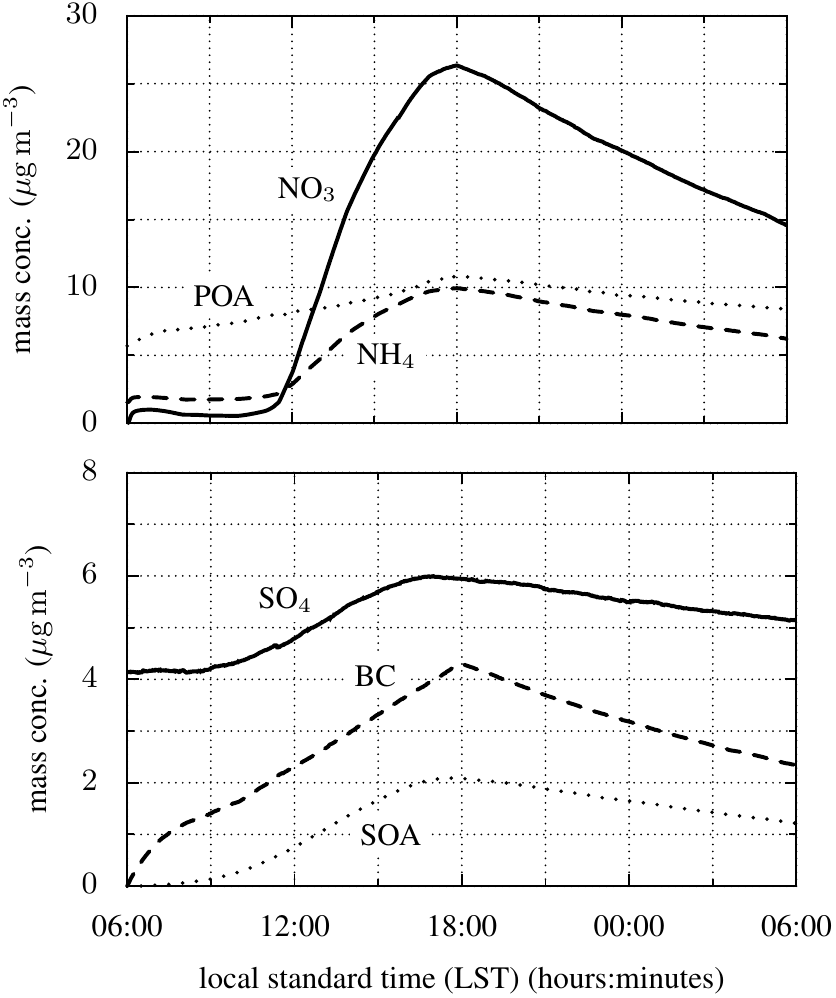}
  \end{center}
  \caption{\label{fig:aging_aero_time_species} Time series of mass
    concentrations of selected aerosol species: nitrate ($\rm NO_3$),
    ammonium ($\rm NH_4$), POA, sulfate ($\rm SO_4$), BC, and
    SOA. Particle and gas phase emissions were present from 06:00 to
    18:00~LST.}
\end{figure}

Figure~\ref{fig:aging_aero_time_species} shows time series of the bulk
aerosol mass concentrations as they result from this urban plume
scenario. We observed a pronounced production of ammonium nitrate,
reaching nitrate mass concentration of up to $26\rm \, \mu g\,m^{-3}$
and ammonium mass concentration of $10\rm \, \mu g\,m^{-3}$ in the
late afternoon. Sulfate mass concentrations increased from $4.1\rm \,
\mu g\,m^{-3}$ to $6.0\rm \, \mu g\,m^{-3}$ due to condensation of
photochemically produced sulfuric acid. POA and BC were directly
emitted (with a temporally constant rate) and accumulated to $11\rm \,
\mu g\,m^{-3}$ and $4.3\rm \, \mu g\,m^{-3}$, respectively, until
18:00~LST when the emissions stopped. After 18:00~LST the mass
concentrations declined due to dilution, especially nitrate and BC for
which the background mass concentration were zero.

\subsection{Characterizing mixing state}
To characterize the mixing state and to discuss the
composition of a particle, we refer to the BC mass fractions as
\begin{align}
\label{eqn:w_defn}
w_{\rm BC,dry} &= \frac{\mu_{\rm BC}}{\mu_{\rm dry}}
\end{align}
where $\mu_{\rm BC}$ is the mass of BC in a given particle and
$\mu_{\rm dry}$ is the total dry mass.

Based on this quantity, we then define a two-dimensional number
concentration that is a function of both particle composition and
diameter. The two-dimensional cumulative number distribution $N_{\rm
  BC, dry}(w, D)$ is the number of particles per volume that have a
diameter less than $D$ and a BC mass fraction of less than $w$. The
top panels in Figure~\ref{fig:aging_aero_2d_bc_ss} show the
corresponding two-dimensional distributions, normalized with the
respective total number concentrations, after 1~hour and
after 24~hours of simulation time. Since even at the time of emission
no particles were pure BC, particles were not present at $w_{{\rm
    BC},{\rm dry}} = 100\%$. Fresh emissions from diesel vehicles
($w_{{\rm BC},{\rm dry}} = 70\%$) and gasoline vehicles ($w_{{\rm
    BC},{\rm dry}} = 20\%$) appear as horizontal lines since particles
in one emission category were all emitted with the same
composition. At $w_{{\rm BC},{\rm dry}} = 0\%$ all the particles
appear that do not contain any BC (i.e. background particles and
particles from meat cooking emissions that have not undergone
coagulation with particles containing BC). After 1~hour (07:00~LST) a
small number of particles between these three classes indicate the
occurrence of coagulation. Comparing this result to the result for the
end of the simulation, we note that at the end of the simulation
particles with diameter below $D = 0.03 \rm \, \mu m$ were heavily
depleted due to coagulation. A continuum of mixing states formed
between the extreme mixing states of $w_{{\rm BC},{\rm dry}} = 0\%$
and $w_{{\rm BC},{\rm dry}} = 70\%$.

\subsection{Calculating CCN activity}

Given that we track the composition evolution of each individual
particle throughout the simulation, we can calculate the critical
supersaturation $S_{{\rm c}}$ that the particle needs in order to
activate. We use the concept of a dimensionless hygroscopicity
parameter suggested by \citet{Ghan2001} or \citet{Petters2007}. In
\citet{Petters2007} this parameter is denoted by $\kappa$, and we
adopt their notation for the remainder of the paper. This concept has
the advantage that results from laboratory measurements can be used to
quantify the hygroscopicity of complex compounds for which $\kappa$
values cannot be calculated in a straightforward manner. The overall
$\kappa$ for a particle is the volume-weighted average of the $\kappa$
values of the constituent species. This requires the assignment of
individual $\kappa$ values for each aerosol component in MOSAIC.

\citet{Petters2007} compiled a table (Table~1 in their paper) with
$\kappa$ values for a variety of inorganic and organic species based
on recent laboratory measurements or on thermodynamic model
calculations. For $\rm (NH_4)_2SO_4$ and $\rm NH_4NO_3$ they report
$\kappa$ values of 0.61 and 0.67, based on calculations by
\citet{Clegg1998} and measurements by \citet{Svenningsson2006},
respectively. Based on this we assume $\kappa = 0.65$ for all salts
formed from the $\rm NH_4^+$-$\rm SO_4^{2-}$-$\rm NO_3^-$ system. For all
MOSAIC model species that represent SOA we assume $\kappa = 0.1$,
based on measurements by \citet{Prenni2007}. Following
\citet{Petters2006} we assume $\kappa=0.001$ for POA and $\kappa=0$
for BC. The critical supersaturation $S_{\rm c}$ for a particle of
diameter $D$ and volume-weighted hygroscopicity parameter $\kappa$ is
then given by
\begin{equation}\label{eqn:Sc}
S_{\rm c} = \frac{C}{\sqrt{\kappa D^3}},
\end{equation}
where
\begin{equation}\label{eqn:CA}
C = \sqrt{\frac{4A^3}{27}} \qquad {\rm and} \qquad
A = \frac{4\sigma_wm_w}{R^*T\rho_w},
\end{equation}
with $\sigma_w$ being the surface tension of water, $m_w$ the
molecular weight of water, $R^*$ the universal gas constant, $\rho_w$
the water density, and $T$ the temperature.

\begin{figure}
  \begin{center}
    \includegraphics{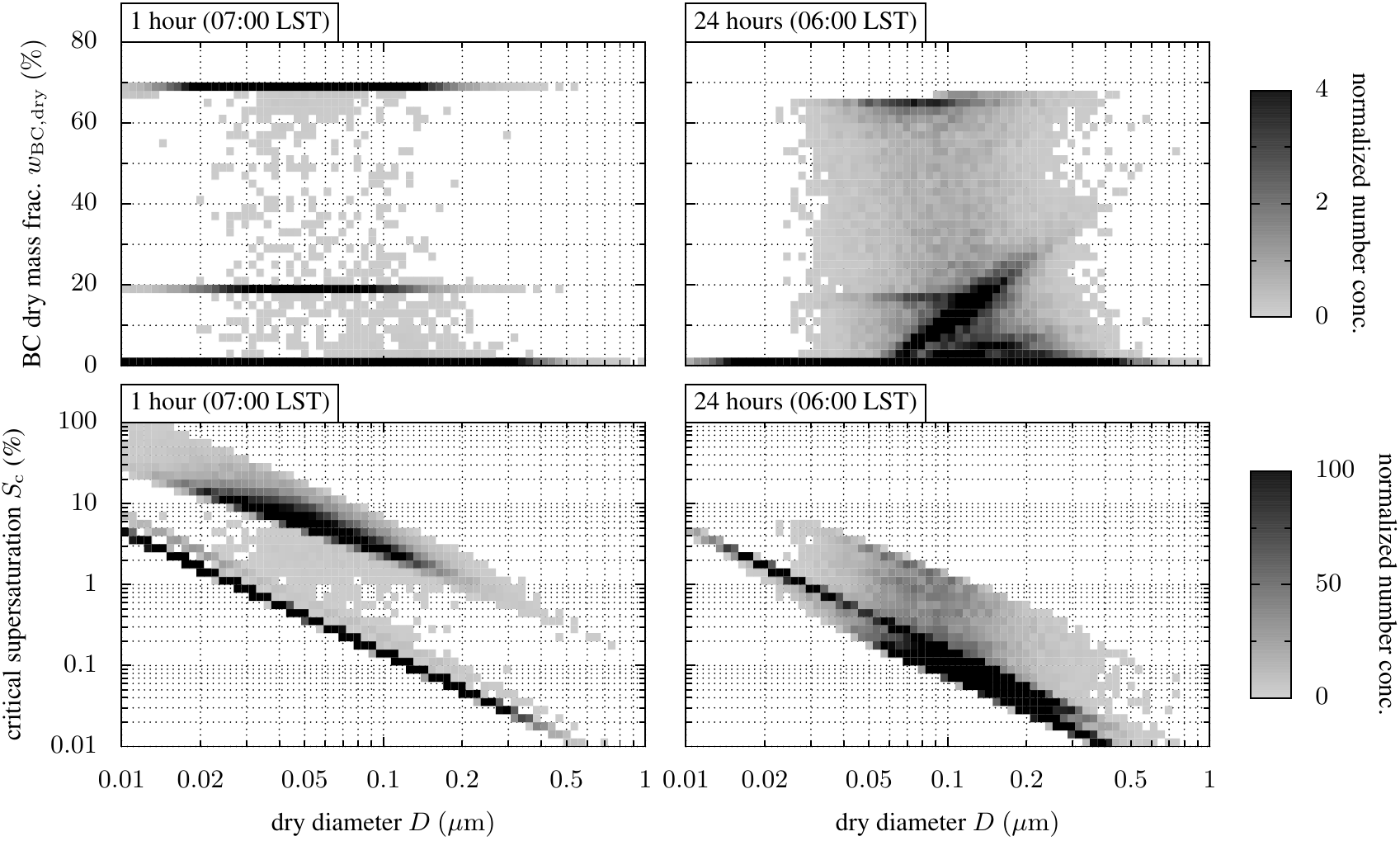}
  \end{center}
  \caption{\label{fig:aging_aero_2d_bc_ss} Normalized two-dimensional
    number distributions after 1~hour (07:00~LST) and 24~hours
    (06:00~LST the next day) of simulation. The top panels show the
    normalized value of the two-dimensional distribution $\partial^2
    N_{\rm BC, dry}(D,w) / (\partial \log_{10} D \ \partial w)$ with
    respect to diameter $D$ and BC dry mass fraction $w_{\rm BC,dry}$,
    while the bottom panels show the normalized value of the
    two-dimensional distribution $\partial^2 N_{\rm S}(D,S_{\rm c}) /
    (\partial \log_{10} D \ \partial \log_{10} S_{\rm c})$ with respect to
    diameter $D$ and critical supersaturation $S_{\rm c}$.}
\end{figure}

Similarly to the use of $w_{\rm BC, dry}$ above, we can use $S_{\rm
  c}$ to define a two-dimensional cumulative number distribution
$N_{\rm S}(D,S_{\rm c})$ in terms of size and critical
supersaturation. The bottom panels in
Figure~\ref{fig:aging_aero_2d_bc_ss} show examples of the
corresponding two-dimensional distributions after 1~hour (left) and
after 24~hours (right) of simulation. While freshly emitted diesel,
gasoline and meat cooking particles differ in their BC and POA mass
fractions, they are very similar in their hygroscopicity with initial
$\kappa$ values close to zero. After 1~hour they are visible as the
dark band of high number concentrations at high $S_c$
values. Separated from this we see another dark band representing the
most hygroscopic particles, consisting of wet background
particles. They contain the largest fraction of inorganic mass
(ammonium, sulfate, and nitrate), hence their critical supersaturation
is lowest at a given size compared to the other particle classes.

Directly slightly above the most hygroscopic band at 1~hour is a
weaker band, which represents the dry background particles. Because
they are dry and the vapor pressure of $\rm HNO_3$ is still low,
nitrate formation does not occur on these particles. The coexistence
of wet and dry particles can be explained by the fact that the
relative humidity falls below 85\% at 06:42~LST, which is the
deliquescence point of the inorganic mixture of ammonium, sulfate, and
nitrate. Particles that exist before that time contain water and take
up nitrate. They stay wet throughout the whole day as a result of the
hysteresis of particle deliquescence and crystallization. Particles
that are emitted after 06:42~LST do not contain water and take up
nitrate only much later. Hence after 1~hour of simulation, at a given
size the fraction of highly hygroscopic inorganics is higher for the
wet particles, which results in a higher $\kappa$ value and lower
critical supersaturation $S_c$.

The individual bands are not completely separated at 1~hour, but the
regions in between have started to fill out.  The reason for this is
the occurance of coagulation, which produces particles of intermediate
composition and hence corresponding intermediate $S_{\rm c}$
values. After 24~hours the population as a whole has moved to lower
critical supersaturations, and the distribution with respect to
$S_{\rm c}$ has become more continuous. Given a certain size, the
critical supersaturation ranges over about one order of magnitude.

\begin{figure}
  \begin{center}
    \includegraphics{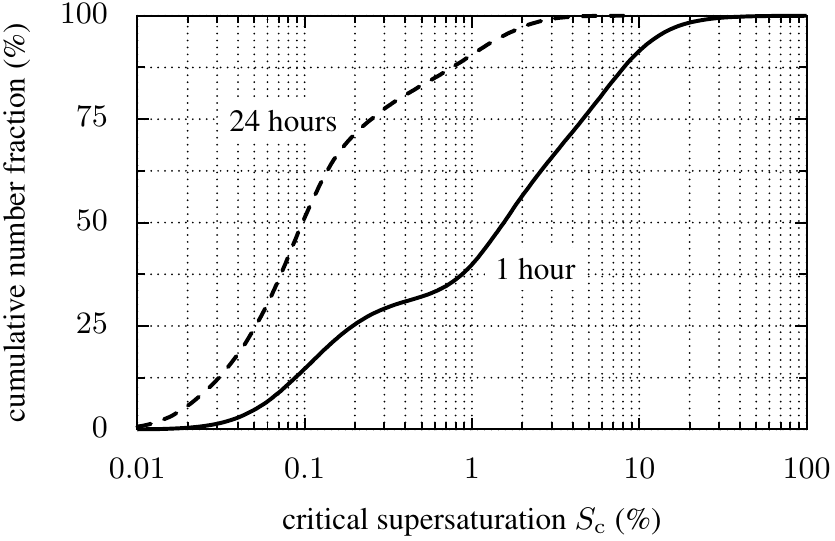}
  \end{center}
  \caption{\label{fig:aging_ccn_spectra} Cloud condensation nuclei
    spectra after 1~hour (07:00~LST) and 24~hours (06:00~LST the next
    day) of simulation based on the simulation results shown in
    Figure~\ref{fig:aging_aero_2d_bc_ss}.}
\end{figure}

Figure~\ref{fig:aging_ccn_spectra} shows CCN properties as more
traditional CCN spectra. This representation is the one-dimensional
projection of the bottom panels of
Figure~\ref{fig:aging_aero_2d_bc_ss} onto the critical supersaturation
axis, plotted as a cumulative distribution. The change in CCN
properties over the course of 24~hours is obvious. After 1~hour a
supersaturation of $S=1.5\%$ is necessary to activate $50\%$ of the
particles by number. This required supersaturation decreases to
$S=0.1\%$ after 24~hours. In the following section we use the results
of this urban plume scenario as a basis for estimating the aging
time-scales.

\section{First-order Models of Aging}
\label{sec:aging_model}

In this section we describe the first-order model of black carbon
aging to which we fit the particle-resolved data simulated with
PartMC-MOSAIC, in order to determine the aging time-scale. We
emphasize that we do not actually simulate using the first-order
models presented in this section. Such first-order systems are
frequently used, however, to model the conversion from hydrophobic to
hydrophilic black carbon with the single parameter of aging rate or
its inverse, the aging time-scale $\tau$ \citep{Croft2005}. More
specifically, the aging time-scale represents the time-scale on which
particles that would initially not activate turn into particles that
can be activated, given a certain chosen supersaturation
threshold. Budget equations for the fresh and aged populations can be
formulated in terms of either number or mass.  A number based aging
time-scale is relevant for aerosol indirect forcing as the cloud
optical properties depend on the cloud droplet number distribution. On
the other hand, a mass aging time-scale is relevant for in-cloud
scavenging and wet removal of BC mass. In the following we will
present results for both number- and mass-based aging time-scales.

At time $t$, the total number concentration $N_{\rm BC}(t)$ of
BC-particles is the sum of the number concentration of BC-containing
fresh particles $N_{\rm f}(t)$ and the number concentration of
BC-containing aged particles $N_{\rm a}(t)$. We define analogously the
total BC mass concentration $M_{\rm BC}(t)$, the BC mass concentration
in fresh BC-containing particles $M_{\rm f}(t)$, and the BC mass
concentration in aged BC-containing particles $M_{\rm a}(t)$. The aged
and fresh populations are separated by applying an aging criterion, in
our case activation at a certain supersaturation threshold $S_{\rm
  c}$. Fresh particles are those with critical supersaturation above
the threshold value, while aged particles have critical
supersaturations below the threshold.

In the PartMC model we explicitly track a finite number of particles
in a computational volume $V$. The number of fresh and aged
BC-containing particles in the volume $V(t_k)$ at time $t_k$ is
denoted by $n_{\rm f}(t_k)$ and $n_{\rm a}(t_k)$,
respectively. Similarly, $m_{\rm f}(t_k)$ and $m_{\rm a}(t_k)$ are
respectively the total mass of BC in fresh and aged BC-containing
particles in $V(t_k)$. The number and mass concentrations of fresh
BC-containing particles in $V(t_k)$ are then given by
\begin{equation}
  \label{eqn:cont_disc_relation}
  \begin{aligned}
    N_{\rm f}(t_k) &= \frac{n_{\rm f}(t_k)}{V(t_k)} &\qquad
    M_{\rm f}(t_k) &= \frac{m_{\rm f}(t_k)}{V(t_k)} \\
    N_{\rm a}(t_k) &= \frac{n_{\rm a}(t_k)}{V(t_k)} &
    M_{\rm a}(t_k) &= \frac{m_{\rm a}(t_k)}{V(t_k)}.
  \end{aligned}
\end{equation}

The fresh and aged number and mass concentrations can change due to
emission and dilution, while condensation and coagulation can transfer
number and mass concentration from the fresh to the aged population
and vice versa. Changes in number and mass concentrations also occur
due to temperature and pressure changes. In our model we neglect at
present the impact of heterogeneous reactions on the surface of the
particles, although studies have shown that these also contribute to
the aging process \citep{Rudich2007}. The gain and loss terms for the
fresh and aged BC-containing populations are given in
Table~\ref{tab:term_def}.

To express changes in number and mass for coagulation we consider that
all constituent particles are lost during a coagulation event and the
product of coagulation is a gain of a new
particle. Table~\ref{tab:combinations} shows the overview of all
possible combinations and the resulting terms for each of those
combinations. For example, let us assume that there are four
independent coagulation events within a single timestep: one event
between two fresh BC-containing particles resulting in an aged
particle, two events between fresh and aged BC-containing particles
each resulting in an aged particle, and one event between a fresh
BC-containing particle and a non-BC-containing particle resulting in a
fresh particle. Then we have losses $\Delta n^{\rm coag}_{\rm f \to
  a}(t_{k-1},t_k) = 5$ and $\Delta n^{\rm coag}_{\rm a \to
  a}(t_{k-1},t_k) = 2$, and gains $\Delta n^{\rm coag}_{\rm
  f}(t_{k-1},t_k) = 1$ and $\Delta n^{\rm coag}_{\rm a}(t_{k-1},t_k) =
3$.

Note that for each particle pairing in Table~\ref{tab:combinations}
there can be two outcomes. For example, coagulation of a small fresh
and large aged particle generally produces an aged particle, while
coagulation of large fresh and small aged generally produces a fresh
particle.  When $S_c$, as calculated in
equations~(\ref{eqn:Sc})--(\ref{eqn:CA}), is used as the criterion for
fresh versus aged, it can be shown that coagulation of two aged
particles never produces a fresh particle, and that the coagulation of
two fresh particles with $S_c$ values close to the cutoff value can
produce an aged particle.

Coagulation can result in a net loss of number but must conserve mass.
We thus have
\begin{align}
  \label{eqn:coag_num_cont_ineq}
  \dot{N}^{\rm coag}_{\rm a}(t)
  &\le \dot{N}^{\rm coag}_{\rm f \to a}(t)
  + \dot{N}^{\rm coag}_{\rm a \to a}(t) \\
  \label{eqn:coag_mass_cont_eq}
  \dot{M}^{\rm coag}_{\rm a}(t)
  &= \dot{M}^{\rm coag}_{\rm f \to a}(t)
  + \dot{M}^{\rm coag}_{\rm a \to a}(t) \\
  \label{eqn:coag_num_disc_ineq}
  \Delta n^{\rm coag}_{\rm a}(t_{k-1},t_k)
  &\le \Delta n^{\rm coag}_{\rm f \to a}(t_{k-1},t_k)
  + \Delta n^{\rm coag}_{\rm a \to a}(t_{k-1},t_k) \\
  \label{eqn:coag_mass_disc_eq}
  \Delta m^{\rm coag}_{\rm a}(t_{k-1},t_k)
  &= \Delta m^{\rm coag}_{\rm f \to a}(t_{k-1},t_k)
  + \Delta m^{\rm coag}_{\rm a \to a}(t_{k-1},t_k)
\end{align}
and similarly for coagulation resulting in fresh particles.

The following continuous equations describe the evolution of the
number and mass concentrations of fresh and aged BC-containing
populations:
\begin{align}
  \label{eqn:num_dot_f}
  \frac{dN_{\rm f}(t)}{dt} &=
  \dot{N}^{\rm density}_{\rm f}(t)
  + \dot{N}^{\rm emit}_{\rm f}(t)
  - \dot{N}^{\rm dilution}_{\rm f}(t)
  \\ &\qquad
  + \dot{N}^{\rm cond}_{\rm a \to f}(t)
  + \dot{N}^{\rm coag}_{\rm f}(t)
  - \dot{N}^{\rm coag}_{\rm f \to f}(t)
  - \underbrace{\Bigl(
    \dot{N}^{\rm cond}_{\rm f \to a}(t)
    + \dot{N}^{\rm coag}_{\rm f \to a}(t) \nonumber
    \Bigr)}_{\dot{N}^{\rm aging}(t)}
  \displaybreak[0] \\
  \label{eqn:num_dot_a}
  \frac{dN_{\rm a}(t)}{dt} &=
  \dot{N}^{\rm density}_{\rm a}(t)
  + \dot{N}^{\rm emit}_{\rm a}(t)
  - \dot{N}^{\rm dilution}_{\rm a}(t)
  \\ &\qquad
  + \dot{N}^{\rm cond}_{\rm f \to a}(t)
  + \dot{N}^{\rm coag}_{\rm a}(t)
  - \dot{N}^{\rm coag}_{\rm a \to a}(t)
  - \underbrace{\Bigl(
    \dot{N}^{\rm cond}_{\rm a \to f}(t)
    + \dot{N}^{\rm coag}_{\rm a \to f}(t)
    \Bigr)}_{\dot{N}^{\text{de-aging}}(t)} \nonumber
  \displaybreak[0] \\
  \label{eqn:mass_dot_f}
  \frac{dM_{\rm f}(t)}{dt} &=
  \dot{M}^{\rm density}_{\rm f}(t)
  + \dot{M}^{\rm emit}_{\rm f}(t)
  - \dot{M}^{\rm dilution}_{\rm f}(t)
  \\ &\qquad
  + \underbrace{\Bigl(
    \dot{M}^{\rm cond}_{\rm a \to f}(t)
    + \dot{M}^{\rm coag}_{\rm a \to f}(t)
    \Bigr)}_{\dot{M}^{\rm de-aging}(t)}
  - \underbrace{\Bigl(
    \dot{M}^{\rm cond}_{\rm f \to a}(t)
    + \dot{M}^{\rm coag}_{\rm f \to a}(t)
    \Bigr)}_{\dot{M}^{\rm aging}(t)} \nonumber
  \displaybreak[0] \\
  \label{eqn:mass_dot_a}
  \frac{dM_{\rm a}(t)}{dt} &=
  \dot{M}^{\rm density}_{\rm a}(t)
  + \dot{M}^{\rm emit}_{\rm a}(t)
  - \dot{M}^{\rm dilution}_{\rm a}(t)
  \\ &\qquad
  + \underbrace{\Bigl(
    \dot{M}^{\rm cond}_{\rm f \to a}(t)
    + \dot{M}^{\rm coag}_{\rm f \to a}(t)
  \Bigr)}_{\dot{M}^{\rm aging}(t)}
  - \underbrace{\Bigl(
    \dot{M}^{\rm cond}_{\rm a \to f}(t)
    + \dot{M}^{\rm coag}_{\rm a \to f}(t)
  \Bigr)}_{\dot{M}^{\rm de-aging}(t)} \nonumber
\end{align}
Note that the mass
equations~(\ref{eqn:mass_dot_f})--(\ref{eqn:mass_dot_a}) have been
simplified from forms identical to the number
equations~(\ref{eqn:num_dot_f})--(\ref{eqn:num_dot_a}) by using
equation~(\ref{eqn:coag_mass_cont_eq}) for the conservation of mass
during coagulation. Condensation (or rather evaporation) and
coagulation can in principle also produce a transfer from aged to
fresh particles. This is reflected by the terms denoted as
$\dot{N}^{\rm de-aging}(t)$ and $\dot{M}^{\rm de-aging}(t)$ above.

The aging time-scales $\tau_{\rm N}$ and $\tau_{\rm M}$ for number and
mass are then determined by the first-order models:
\begin{align}
  \label{eqn:cont_aging_model_N}
  \dot{N}^{\rm aging}(t) &= \frac{1}{\tau_{\rm N}(t)} N_{\rm f}(t) \\
  \label{eqn:cont_aging_model_M}
  \dot{M}^{\rm aging}(t) &= \frac{1}{\tau_{\rm M}(t)} M_{\rm f}(t)
\end{align}

The discrete versions of the balance
equations~(\ref{eqn:num_dot_f})--(\ref{eqn:mass_dot_a}) are:
\begin{align}
  \label{eqn:n_dot_disc_f}
  n_{\rm f}(t_k) - n_{\rm f}(t_{k-1}) &=
  \Delta n^{\rm emit}_{\rm f}(t_{k-1},t_k)
  - \Delta n^{\rm dilution}_{\rm f}(t_{k-1},t_k)
   \\ &\qquad \nonumber
  + \Delta n^{\rm cond}_{\rm a \to f}(t_{k-1},t_k)
  + \Delta n^{\rm coag}_{\rm f}(t_{k-1},t_k)
  - \Delta n^{\rm coag}_{\rm f \to f}(t_{k-1},t_k)
  \\ &\qquad \nonumber
  - \underbrace{\Bigl(
    \Delta n^{\rm cond}_{\rm f \to a}(t_{k-1},t_k)
    + \Delta n^{\rm coag}_{\rm f \to a}(t_{k-1},t_k)
  \Bigr)}_{\Delta n^{\rm aging}(t_{k-1},t_k)} \nonumber
  \displaybreak[0] \\
  \label{eqn:n_dot_disc_a}
  n_{\rm a}(t_k) - n_{\rm a}(t_{k-1}) &=
  \Delta n^{\rm emit}_{\rm a}(t_{k-1},t_k)
  - \Delta n^{\rm dilution}_{\rm a}(t_{k-1},t_k)
   \\ &\qquad \nonumber
  + \Delta n^{\rm cond}_{\rm f \to a}(t_{k-1},t_k)
  + \Delta n^{\rm coag}_{\rm a}(t_{k-1},t_k)
  - \Delta n^{\rm coag}_{\rm a \to a}(t_{k-1},t_k)
  \\ &\qquad 
  - \underbrace{\Bigl(
    \Delta n^{\rm cond}_{\rm a \to f}(t_{k-1},t_k)
    + \Delta n^{\rm coag}_{\rm a \to f}(t_{k-1},t_k)
    \Bigr)}_{\Delta n^{\rm de-aging}(t_{k-1},t_k)}
  \nonumber
  \displaybreak[0] \\
  \label{eqn:m_dot_disc_f}
  m_{\rm f}(t_k) - m_{\rm f}(t_{k-1})
  &=
  \Delta m^{\rm emit}_{\rm f}(t_{k-1},t_k)
  - \Delta m^{\rm dilution}_{\rm f}(t_{k-1},t_k)
  \\ &\qquad
  + \underbrace{\Bigl(
    \Delta m^{\rm cond}_{\rm a \to f}(t_{k-1},t_k)
    + \Delta m^{\rm coag}_{\rm a \to f}(t_{k-1},t_k)
    \Bigr)}_{\Delta m^{\rm de-aging}(t_{k-1},t_k)}
  \nonumber
  \\ &\qquad
  - \underbrace{\Bigl(
    \Delta m^{\rm cond}_{\rm f \to a}(t_{k-1},t_k)
    + \Delta m^{\rm coag}_{\rm f \to a}(t_{k-1},t_k)
    \Bigr)}_{\Delta m^{\rm aging}(t_{k-1},t_k)}
  \nonumber
  \displaybreak[0] \\
  \label{eqn:m_dot_disc_a}
  m_{\rm a}(t_k) - m_{\rm a}(t_{k-1})
  &=
  \Delta m^{\rm emit}_{\rm a}(t_{k-1},t_k)
  - \Delta m^{\rm dilution}_{\rm a}(t_{k-1},t_k)
  \\ &\qquad
  + \underbrace{\Bigl(
    \Delta m^{\rm cond}_{\rm f \to a}(t_{k-1},t_k)
    + \Delta m^{\rm coag}_{\rm f \to a}(t_{k-1},t_k)
    \Bigr)}_{\Delta m^{\rm aging}(t_{k-1},t_k)} \nonumber
  \\ &\qquad
  - \underbrace{\Bigl(
    \Delta m^{\rm cond}_{\rm a \to f}(t_{k-1},t_k)
    + \Delta m^{\rm coag}_{\rm a \to f}(t_{k-1},t_k)
    \Bigr)}_{\Delta m^{\rm de-aging}(t_{k-1},t_k)} \nonumber
\end{align}
Here equation~(\ref{eqn:coag_mass_disc_eq}) for the discrete
conservation of mass during coagulation was used to simplify the mass
equations, as in the continuous case.

By comparing the continuous
equations~(\ref{eqn:num_dot_f})--~(\ref{eqn:mass_dot_a}) to the
discrete equations~(\ref{eqn:n_dot_disc_f})--~(\ref{eqn:m_dot_disc_a})
using the relationships~(\ref{eqn:cont_disc_relation}) we see that the
continuous aging terms can be approximated by:
\begin{align}
  \dot{N}^{\rm aging}(t_k) &\approx
  \frac{
    \Delta n^{\rm aging}(t_{k-1},t_k)
  }{(t_k - t_{k-1}) V(t_k)} \displaybreak[0] \\
  \dot{M}^{\rm aging}(t_k) &\approx
  \frac{
    \Delta m^{\rm aging}(t_{k-1},t_k)
  }{(t_k - t_{k-1}) V(t_k)},
\end{align}
where the use of $V$ at time $t_k$ is because the computational volume
$V$ is updated first within each timestep in the PartMC algorithm
\citep[Figure~1]{Riemer2009}.

From equations~(\ref{eqn:cont_aging_model_N})
and~(\ref{eqn:cont_aging_model_M}) and the
relationships~(\ref{eqn:cont_disc_relation}), the aging time-scales
$\tau_{\rm N}$ and $\tau_{\rm M}$ are then approximated by:
\begin{align}\label{eqn:tau_N}
  \tau_{\rm N}(t_k) &\approx
  \frac{(t_k - t_{k-1}) n_{\rm f}(t_k)}{
    \Delta n^{\rm aging}(t_{k-1},t_k)
  } \displaybreak[0] \\
  \tau_{\rm M}(t_k) &\approx
  \frac{(t_k - t_{k-1}) m_{\rm f}(t_k)}{
    \Delta m^{\rm aging}(t_{k-1},t_k)
  }. \label{eqn:tau_M}
\end{align}

For the analysis in Section~\ref{sec:results} we additionally define
the aging time-scale $\tau_{\rm N}^{\rm cond}$ ignoring the impact of
coagulation, an average time-scale $\tau_{\rm N,day}$ during the day,
and an average time-scale $\tau_{\rm N,night}$ during the night, given
by:
\begin{align}
  \label{eqn:tau_N_cond}
  \tau_{\rm N}^{\rm cond}(t_k) &\approx
  \frac{(t_k - t_{k-1}) n_{\rm f}(t_k)}{
    \Delta n^{\rm cond}_{\rm f \to a}(t_{k-1},t_k)} \\
  \label{eqn:tau_N_day}
  (\tau_{\rm N,day})^{-1}
  &= \frac{1}{3 \, {\rm hours}} \int_{\rm 12:00\, LST}^{\rm 15:00\, LST}
  \bigl(\tau_{\rm N}(t)\bigr)^{-1} \, dt \\
  \label{eqn:tau_N_night}
  (\tau_{\rm N,night})^{-1}
  &= \frac{1}{10 \, {\rm hours}} \int_{\rm 18:00\, LST}^{\rm 04:00\, LST}
  \bigl(\tau_{\rm N}(t)\bigr)^{-1} \, dt, 
\end{align}
and similarly for $\tau_{\rm M}^{\rm cond}$, $\tau_{\rm M,day}$,
$\tau_{\rm M,night}$, $\tau_{\rm N,day}^{\rm cond}$, and so on.

\section{Results}
\label{sec:results}

\begin{figure}
  \begin{center}
    \includegraphics{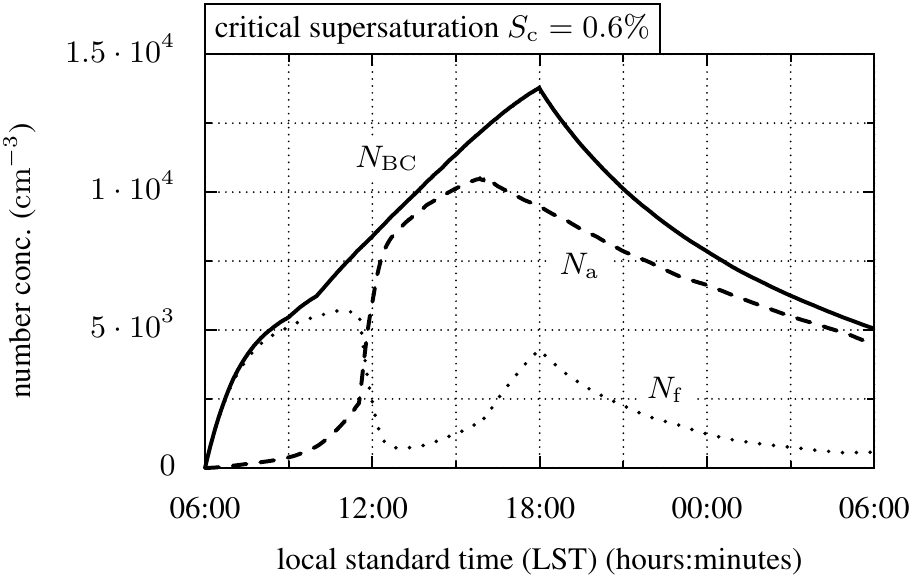}
  \end{center}
  \caption{\label{fig:aging_aero_time_totals} Time series of total
    number concentration $N_{\rm BC}$ of BC-containing particles and
    number concentrations $N_{\rm f}$ and $N_{\rm a}$ for fresh and
    aged BC-containing particles, respectively. The critical
    supersaturation separating fresh and aged particles is set to
    $S_{\rm c}=0.6\%$.}
\end{figure}

In this section we show results for supersaturation thresholds ranging
between $S_{\rm c} = 0.1\%$ and $S_{\rm c} = 1.0\%$, as these are
typically achieved in updrafts in stratus and cumulus clouds
\citep{Warner1968}. To limit the number of figures we use $S_{\rm
  c}=0.6\%$ as a base case. Figure~\ref{fig:aging_aero_time_totals}
shows the time series for $N_{\rm f}$, $N_{\rm a}$, and $N_{\rm
  BC}=N_{\rm f}+N_{\rm a}$ for $S_{\rm c}=0.6\%$.  The total number
concentration $N_{\rm BC}$ of BC-containing particles increased until
18:00~LST due to the emission of particles. After the emissions
stopped, $N_{\rm BC}$ decreased as a result of continued dilution and
coagulation. The time series for $N_{\rm f}$ and $N_{\rm a}$ show that
both increased in the morning hours. A pronounced transfer from fresh
to aged occurred after 11:30~LST, the time when nitrate formation
started taking place on the dry particles (compare
Figure~\ref{fig:aging_aero_time_species}). This process efficiently
contributed to the conversion of fresh particles to aged particles,
which was reflected by a short aging time-scale.

\begin{figure}
  \begin{center}
    \includegraphics{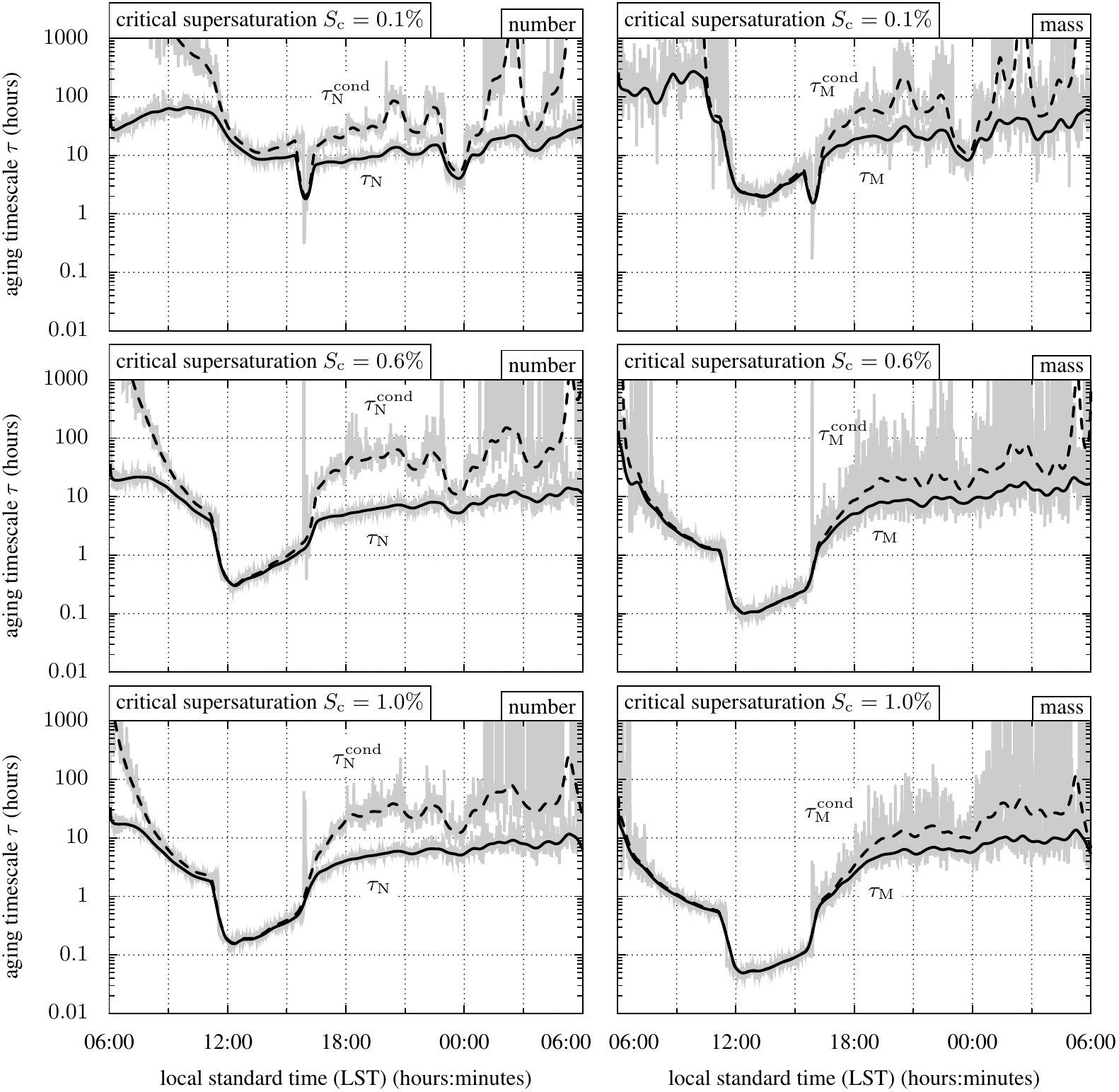}
  \end{center}
  \caption{\label{fig:aging_aero_tau} Comparison of aging time-scales
    based on number (left) and mass (right), showing both the
    time-scales due to condensation and coagulation ($\tau_{\rm N}$
    and $\tau_{\rm M}$) and the time-scales due to condensation alone
    ($\tau_{\rm N}^{\rm cond}$ and $\tau_{\rm M}^{\rm cond}$). The top
    panels have the critical supersaturation separating fresh from
    aged particles set to $S_{\rm c}=0.1\%$, while the middle panels
    have $S_{\rm c}=0.6\%$, and the bottom panels have $S_{\rm c}=
    1.0\%$.}
\end{figure}

The left column of Figure~\ref{fig:aging_aero_tau} compares the aging
time-scales $\tau_{\rm N}$ and $\tau_{\rm N}^{\rm cond}$ computed
according to equations~(\ref{eqn:tau_N}) and~(\ref{eqn:tau_N_cond})
for different supersaturation thresholds. The top, middle and bottom
panels shows the results for supersaturation thresholds $S_{\rm
  c}=0.1\%$, $S_{\rm c}=0.6\%$, and $S_{\rm c}=1\%$, respectively. The
grey shading is the raw data, and the black lines are results computed
from smoothing the transfer rates with a Hann filter with a window
width of 1~hour.

The solid lines represent $\tau_{\rm N}$, including the contributions
due to both coagulation and condensation according to
equation~(\ref{eqn:tau_N}). For $S_{\rm c}=0.6\%$, $\tau_{\rm N}$
started off in the morning with values of $\tau_{\rm N} \approx
20$~hours. It decreased sharply after 11:00~LST, which is the time
when photochemistry was at its peak, and nitrate formation was most
pronounced, hence leading to a fast aging process. Between 12:00~LST
and 16:00~LST, $\tau_{\rm N}$ was less than 1~hour, and reached values
as low as 0.2~hours. After 16:00~LST, as photochemistry slowed down,
$\tau_{\rm N}$ increased again, reaching a plateau of $\tau_{\rm N}
\approx 10$~hours during the evening and night.

The broken lines represent $\tau_{\rm N}^{\rm cond}$ as defined in
equation~(\ref{eqn:tau_N_cond}), i.e. the contribution due to
coagulation is ignored. There was only a small difference between
$\tau_{\rm N}$ and $\tau_{\rm N}^{\rm cond}$ during midday and early
afternoon when condensation was operating very effectively. However
during morning, afternoon and night, neglecting coagulation lead to
larger time-scales (up to one order of magnitude).

A similar pattern was found for $S_{\rm c}=1\%$, shown in the bottom
panel. For $S_{\rm c}=0.1\%$ we generally obtained larger
time-scales. During the morning, $\tau_{\rm N}$ was about 50~hours and
decreased to 10~hours in the early afternoon. There was a short period
around 16:00~LST when $\tau_{\rm N}$ dropped to 2~hours. Obviously, at
$S_{\rm c}=0.1\%$, even after the growth due to condensation of
ammonium nitrate, the particles were still too small to be activated
to the same extent as seen for the larger supersaturations. During the
following night $\tau_{\rm N}$ was around 10--20~hours. For this low
supersaturation the contrast between day and night was not as
pronounced as for $S_{\rm c}=0.6\%$ or $S_{\rm c}=1\%$.

Qualitatively, the temporal evolution of $\tau_{\rm M}$ (right column)
was similar to the number-based result $\tau_{\rm N}$. However the
day/night contrast was more pronounced for the case with $S_{\rm
  c}=0.1\%$, and the time-scales based on mass during the day were
lower than the ones based on number.

\begin{figure}
  \begin{center}
    \includegraphics{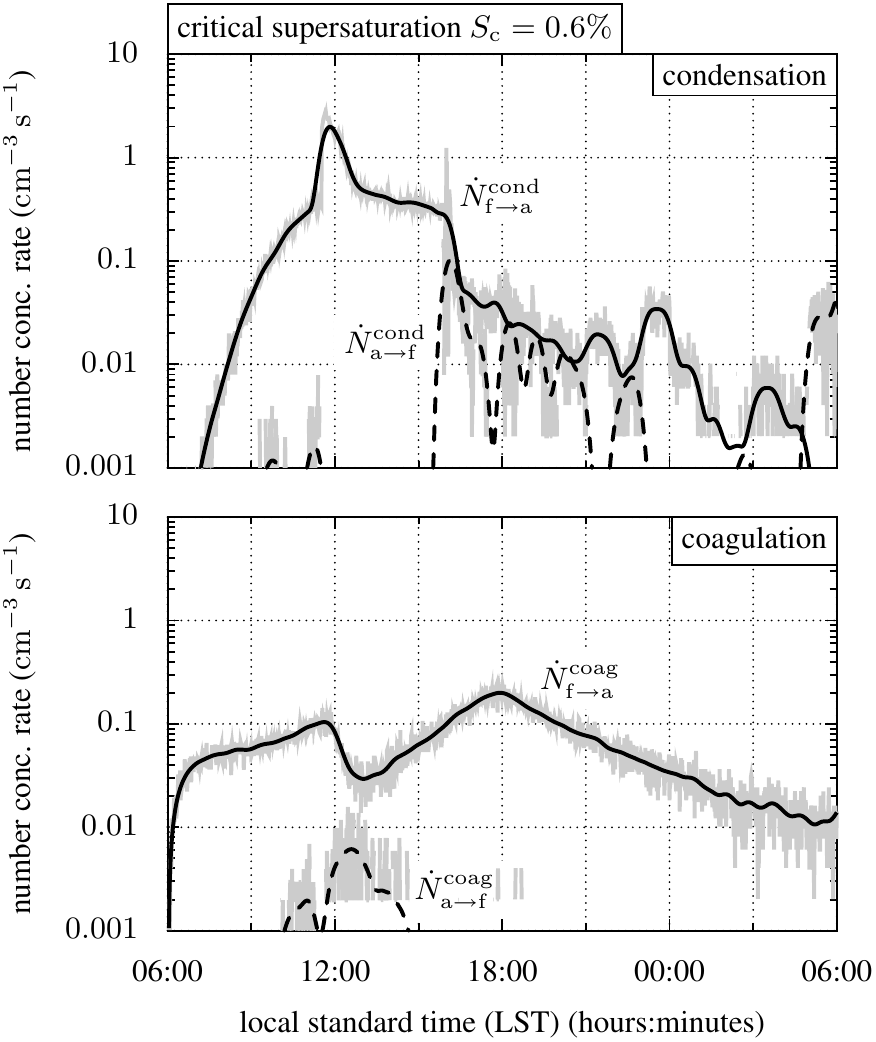}
  \end{center}
  \caption{\label{fig:aging_aero_time_transfers} Transfer rates due to
    condensation (top) and coagulation (bottom). The critical
    supersaturation separating fresh and aged particles is set to
    $S_{\rm c}=0.6\%$. The notation is according to
    Table~\ref{tab:term_def}.}
\end{figure}

Figure~\ref{fig:aging_aero_time_transfers} shows the individual
transfer terms of BC-containing particles for the case $S_{\rm
  c}=0.6\%$.  The transfer rate $\dot N_{\rm a \to f}^{\rm coag}$ due
to coagulation from aged to fresh was very small throughout the whole
day, remaining below $0.01$~$\rm cm ^{-3}\, s^{-1}$. The transfer rate
$\dot N_{\rm f \to a}^{\rm coag}$ due to coagulation from fresh to
aged followed the time series of $N_{\rm f}$. The minimum of $N_{\rm
  f}$ during the early afternoon was reflected in a minimum of the
transfer rate $\dot N_{\rm f \to a}^{\rm coag}$. The transfer rate
$\dot N_{\rm f \to a}^{\rm cond}$ due to condensation from fresh to
aged was large between 11:30 and 15:00~LST, consistent with the
decrease of $\tau_{\rm N}$ in Figure~\ref{fig:aging_aero_tau}.
Lastly, there was a non-zero transfer $\dot N_{\rm a \to f}^{\rm
  cond}$ from aged to fresh due to condensation, which was larger than
$\dot N_{\rm f \to a}^{\rm cond}$ towards the end of the
simulation. This was related to a shrinking of the particles due to
decreasing relative humidity (compare Figure~\ref{fig:aging_env}).

\begin{figure}
  \begin{center}
    \includegraphics[width=14cm]{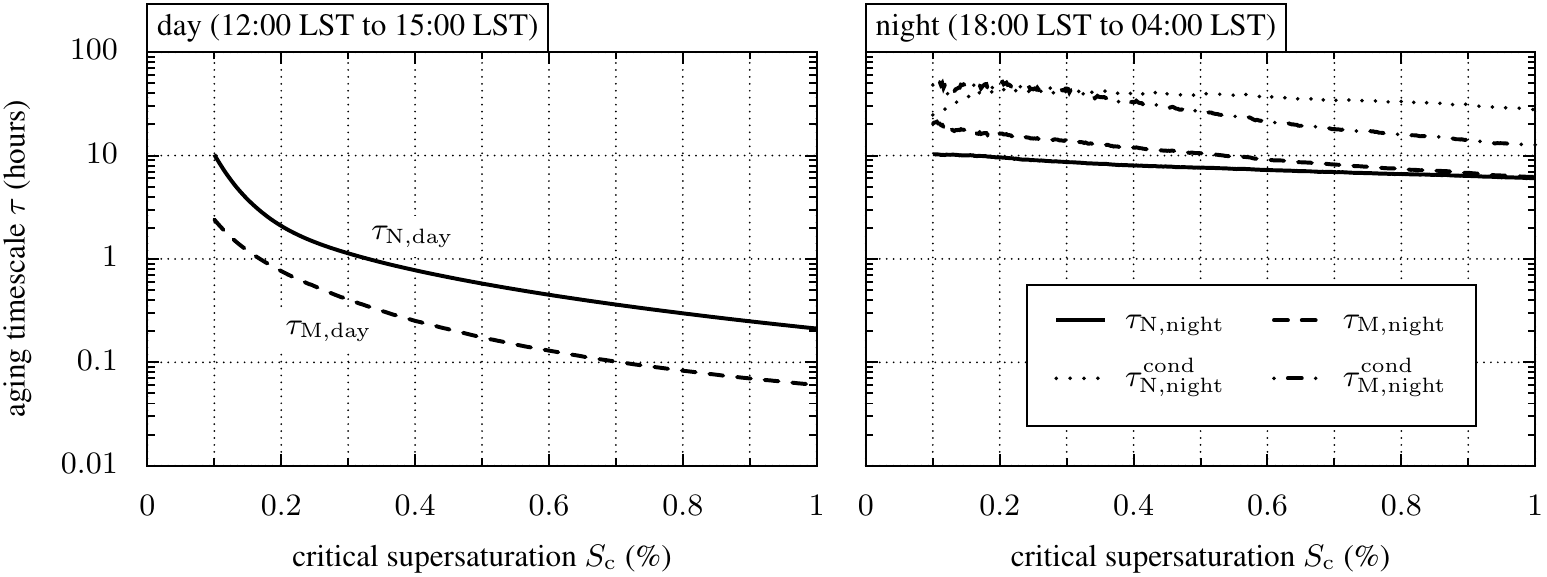}
  \end{center}
  \caption{\label{fig:aging_tau_day_night} Day (left) and night
    (right) averages of the aging time-scale, as defined in
    equations~(\ref{eqn:tau_N_cond})--(\ref{eqn:tau_N_night}). As can
    be see from Figure~\ref{fig:aging_aero_tau}, during the day the
    condensation-induced aging dominates, so $\tau^{\rm cond}_{\rm
      N,day}$ is almost indistinguishable from $\tau_{\rm N,day}$, and
    similarly for mass.}
\end{figure}

Figure~\ref{fig:aging_tau_day_night} summarizes the results for the
different aging time-scale definitions. We calculated $\tau_{\rm
  N,day}$ and $\tau_{\rm N,night}$ according to
equation~(\ref{eqn:tau_N_cond}) for several hundred different
supersaturation thresholds between $S_{\rm c} = 0.1\%$ and $S_{\rm c}
= 1\%$. We also distinguished between the definition of $\tau$
including the transfer due to coagulation and condensation (solid
lines), and including only the transfer due to condensation,
$\tau_{\rm N,day}^{\rm cond}$ and $\tau_{\rm N,night}^{\rm cond}$
(broken lines).

For this particular urban plume scenario the following general
features emerge: during the day, condensation of semi-volatile
species, in our case especially ammonium nitrate, was the dominant
process for aging. The time-scales based on number were larger than
the time-scales based on mass during the day, by roughly a factor of
five. The aging time-scales had a strong dependence on supersaturation
threshold. For $S_{\rm c}=0.1\%$, $\tau_{\rm N,day}$ was 10~hours,
whereas for $S_{\rm c}=1\%$, $\tau_{\rm N,day}$ was only
0.2~hours. During the night condensation was limited, hence
coagulation was the dominant aging process. For low supersaturation
thresholds, the time-scales based on number were smaller than the
time-scales based on mass, but this difference decreased for higher
supersaturation thresholds.

\section{Conclusions}
\label{sec:conclusion}

In this paper we presented a method for explicitly calculating aging
time-scales of black carbon aerosol using particle-resolved model
simulations with PartMC-MOSAIC. We developed number-based and
mass-based aging time-scales using the activation of the particles at
a given supersaturation as a criterion for aging. We applied this
method to an urban plume scenario \citep{Riemer2009} for a range of
supersaturation thresholds between 0.1\% and 1\% and considered
condensation of secondary substances and coagulation as aging
mechanisms. Aging due to heterogeneous processes (chemical aging) were
not included.

For this particular scenario we found a separation into day and night
regimes. During the day the condensation-induced aging dominated, in
particular due to the formation of ammonium nitrate. Therefore the
condensation-only number-based aging time-scale $\tau_{\rm N,
  day}^{\rm cond}$ was almost the same as the total time-scale
$\tau_{\rm N, day}$ and similarly for mass. The daytime aging
number-based time-scale $\tau_{\rm N, day}$ was about 10~hours for a
supersaturation threshold of $S_{\rm c}=0.1 \%$ and decreased to
0.06~hours for a threshold of $S_{\rm c}=1\%$. The daytime mass-based
aging time-scale $\tau_{\rm M, day}$ was about a factor of five lower
than $\tau_{\rm N, day}$ for all supersaturation thresholds.

During the night, the absence of condensable species caused the
number-based aging time-scale $\tau_{\rm N, night}$ to be about one
order of magnitude larger than during the day. Coagulation became
dominant, which was reflected by the fact that the condensation-only
time-scale $\tau_{\rm N, night}^{\rm cond}$ was an order of magnitude
larger than the total time-scale $\tau_{\rm N, night}$. The nighttime
aging time-scales therefore depended on the particle number
concentrations. We suspect that chemical aging would have its largest
impact during periods when condensation is not dominant, i.e. during
the night in our case.

Compared to the time-scales used in global models, which are typically
on the order of 30~hours \citep[e.g.]{Chung2002,Koch2001}, our
time-scales were much shorter, in particular during the day. This
confirmed findings by \citet{Riemer2004} who showed with a completely
different approach that daytime and nighttime aging regimes exists,
and that aging during the day proceeds very rapidly.

However some caveats need to be emphasized: our urban plume scenario
represents only one scenario, with very polluted conditions and fairly
high number concentrations during the night ($N_{\rm BC} \approx
5000$~$\rm cm^{-3}$). For lower number concentrations, we expect the
aging time-scales during the night to increase.

\paragraph*{Acknowledgments.}
Funding for N.~Riemer and M.~West was provided by the National Science
Foundation (NSF) under grant ATM 0739404. Funding for R.~A.~Zaveri and
R.~C.~Easter was provided by the Aerosol-Climate Initiative as part of
the Pacific Northwest National Laboratory (PNNL) Laboratory Directed
Research and Development (LDRD) program. Pacific Northwest National
Laboratory is operated for the U.S. Department of Energy by Battelle
Memorial Institute under contract DE-AC06-76RLO 1830.

\bibliographystyle{elsart-harv}
\bibliography{refs}

\begin{thebibliography}{41}
\expandafter\ifx\csname natexlab\endcsname\relax\def\natexlab#1{#1}\fi
\expandafter\ifx\csname url\endcsname\relax
  \def\url#1{\texttt{#1}}\fi
\expandafter\ifx\csname urlprefix\endcsname\relax\def\urlprefix{URL }\fi

\bibitem[{Andreae and Gelenc\'ser(2006)}]{Andreae2006}
Andreae, M., Gelenc\'ser, A., 2006. Black carbon or brown carbon? {T}he nature
  of light-absorbing carbonaceous aerosols. Atmos. Chem. Phys. 6, 3131--3148.

\bibitem[{Bond et~al.(2004)Bond, Streets, Yarber, Nelson, Woo, and
  Klimont}]{Bond2004}
Bond, T., Streets, D., Yarber, K., Nelson, S., Woo, J., Klimont, Z., 2004. {A
  technology-based global inventory of black and organic carbon emissions from
  combustion}. J. Geophys. Res 109~(D14), D14203.

\bibitem[{Chung and Seinfeld(2002)}]{Chung2002}
Chung, S.~H., Seinfeld, J.~H., 2002. Global distribution and climate forcing of
  carbonaceous aerosols. J. Geophys. Res. 107.

\bibitem[{Clegg et~al.(1998)Clegg, Brimblecombe, and Wexler}]{Clegg1998}
Clegg, S., Brimblecombe, P., Wexler, A., 1998. {Thermodynamic model of the
  system $\rm H^+$-$\rm NH^+_4$-$\rm Na^+$-$\rm SO^{2-}_4$-$\rm NH_3$-$\rm
  Cl^-$-$H_2O$ at 298.15 K}. J. Phys. Chem. A 102~(12), 2155--2171.

\bibitem[{Cooke et~al.(1999)Cooke, Liousse, Cachier, and Feichter}]{Cooke1999}
Cooke, W.~F., Liousse, C., Cachier, H., Feichter, J., 1999. Construction of a
  1$^\circ \times 1^\circ$ fossil fuel emission data set for carbonaceous
  aerosol and implementation and radiative impact in the {ECHAM}4 model. J.
  Geophys. Res. 104, 22137--22162.

\bibitem[{Croft et~al.(2005)Croft, Lohmann, and von Salzen}]{Croft2005}
Croft, B., Lohmann, U., von Salzen, K., 2005. Black carbon aging in the
  {C}anadian {C}entre for {C}limate modelling and analysis atmospheric general
  circulation model. Atmos. Chem. Phys. 5, 1931--1949,
  sRef-ID:1680-7324/acp/2005-5-1931.

\bibitem[{Cubison et~al.(2008)Cubison, Ervens, Feingold, Docherty, Ulbrich,
  Shields, Prather, Hering, and Jimenez}]{Cubison2008}
Cubison, M.~J., Ervens, B., Feingold, G., Docherty, K.~S., Ulbrich, I.~M.,
  Shields, L., Prather, K., Hering, S., Jimenez, J.~L., 2008. The influence of
  chemical composition and mixing state on {L}os {A}ngeles urban aerosol on
  {CCN} number and cloud properties. Atmos. Chem. Phys. Discuss. 8, 5629--5681.

\bibitem[{Eldering and Cass(1996)}]{Eldering1996}
Eldering, A., Cass, G.~R., 1996. Source-oriented model for air pollution
  effects on visibility. J. Geophys. Res. 101, 19343--19369.

\bibitem[{Furutani et~al.(2008)Furutani, Dall’osto, Roberts, and
  Prather}]{Furutani2008}
Furutani, H., Dall’osto, M., Roberts, G., Prather, K., 2008. {Assessment of
  the relative importance of atmospheric aging on {CCN} activity derived from
  field observations}. Atmos. Environ. 42~(13), 3130--3142.

\bibitem[{Ghan et~al.(2001)Ghan, Laulainen, Easter, Wagener, Nemesure, Chapman,
  Zhang, and Leung}]{Ghan2001}
Ghan, S., Laulainen, N., Easter, R., Wagener, R., Nemesure, S., Chapman, E.,
  Zhang, Y., Leung, R., 2001. {Evaluation of aerosol direct radiative forcing
  in MIRAGE}. Journal of Geophysical Research 106~(D6), 5317--5334.

\bibitem[{Horvath and Trier(1993)}]{Horvath1993}
Horvath, H., Trier, A., 1993. A study of the aerosol of {S}antiago de {C}hile
  --- {I}. {L}ight extinction coefficient. Atmos. Environ. 27, 371--384.

\bibitem[{IPCC(2007)}]{IPCC2007}
IPCC, 2007. Climate Change 2007: {T}he physical science basis summary for
  policymakes. {C}ontribution of working group {I} to the fourth assessment
  report of the {I}ntergovernmental {P}anel on {C}limate {C}hange. World
  Meteorological Organization, Geneva, Switzerland.

\bibitem[{Jaenicke(1993)}]{Jaenicke1993}
Jaenicke, R., 1993. Aerosol-Cloud-Climate Interaction. Academic Press, San
  Diego, CA, Ch. Tropospheric aerosols, pp. 1--31.

\bibitem[{Kittelson et~al.(2006{\natexlab{a}})Kittelson, Watts, and
  Johnson}]{Kittelson2006-1}
Kittelson, D., Watts, W., Johnson, J., 2006{\natexlab{a}}. On-road and
  laboratory evaluation of combustion aerosols --- {P}art 1: {S}ummary of
  diesel engine results. Aerosol Sci. 37, 913--930.

\bibitem[{Kittelson et~al.(2006{\natexlab{b}})Kittelson, Watts, Johnson,
  Schauer, and Lawson}]{Kittelson2006-2}
Kittelson, D., Watts, W., Johnson, J., Schauer, J., Lawson, D.,
  2006{\natexlab{b}}. On-road and laboratory evaluation of combustion aerosols
  --- {P}art 2: {S}ummary of spark ignition engine results. Aerosol Sci. 37,
  931--949.

\bibitem[{Kleeman et~al.(2000)Kleeman, Schauer, and Cass}]{Kleeman2000}
Kleeman, M., Schauer, J., Cass, G., 2000. Size and composition distribution of
  fine particulate matter emitted from motor vehicles. Environ. Sci. Technol.
  34, 1132--1142.

\bibitem[{Koch(2001)}]{Koch2001}
Koch, D., 2001. Transport and direct radiative forcing of carbonaceous and
  sulfate aerosols in the {GISS GCM}. J. Geophys. Res. 106, 20311--20332.

\bibitem[{Lohmann et~al.(1999)Lohmann, Feichter, Chuang, and
  Penner}]{Lohmann1999}
Lohmann, U., Feichter, J., Chuang, C.~C., Penner, J.~E., 1999. Prediction of
  the number of cloud droplets in the {ECHAM GCM}. J. Geophys. Res. 104,
  9169--9198.

\bibitem[{McFiggans et~al.(2006)McFiggans, Artaxo, Baltensperger, Coe,
  Facchini, Feingold, Fuzzi, Gysel, Laaksonen, Lohmann, Mentel, Murphy, O'Dowd,
  Snider, and Weingartner}]{McFiggans2006}
McFiggans, G., Artaxo, P., Baltensperger, U., Coe, H., Facchini, M.~C.,
  Feingold, G., Fuzzi, S., Gysel, M., Laaksonen, A., Lohmann, U., Mentel,
  T.~F., Murphy, D.~M., O'Dowd, C.~D., Snider, J.~R., Weingartner, E., 2006.
  The effect of physical and chemical aerosol properties on warm cloud droplet
  activation. Atmos. Chem. Phys. 6, 2593--2649.

\bibitem[{McMurry and Stolzenburg(1989)}]{McMurry1989}
McMurry, P., Stolzenburg, M., 1989. {On the sensitivity of particle size to
  relative humidity for Los Angeles aerosols}. Atmospheric Environment (1967)
  23~(2), 497--507.

\bibitem[{Medalia and Rivin(1982)}]{Medalia1982}
Medalia, A., Rivin, D., 1982. Particulate carbon and other components of soot
  and carbon black. Carbon 20, 481--492.

\bibitem[{Menon et~al.(2002)Menon, Hansen, Nazarenko, and Luo}]{Menon2002}
Menon, S., Hansen, J., Nazarenko, L., Luo, Y.~F., 2002. Climate effects of
  black carbon aerosols in {C}hina and {I}ndia. Science 297, 2250--2253.

\bibitem[{Moffet et~al.(2008)Moffet, Qin, Rebotier, Furutani, and
  Prather}]{Moffet2008}
Moffet, R., Qin, X., Rebotier, T., Furutani, H., Prather, K., 2008. {Chemically
  segregated optical and microphysical properties of ambient aerosols measured
  in a single-particle mass spectrometer}. J. Geophys. Res. 113~(D12), D12213.

\bibitem[{Petters et~al.(2006)Petters, Prenni, Kreidenweis, DeMott, Matsunaga,
  Lim, and Ziemann}]{Petters2006}
Petters, M., Prenni, A., Kreidenweis, S., DeMott, P., Matsunaga, A., Lim, Y.,
  Ziemann, P., 2006. {Chemical aging and the hydrophobic-to-hydrophilic
  conversion of carbonaceous aerosol}. Geophys. Res. Lett 33, L24806.

\bibitem[{Petters and Kreidenweis(2007)}]{Petters2007}
Petters, M.~D., Kreidenweis, S.~M., 2007. A single parameter representation of
  hygroscopic growth and cloud condensation nucleus activity. Atmos. Chem.
  Phys. 7, 1961--1971.

\bibitem[{P\"oschl(2005)}]{Poeschl2005}
P\"oschl, U., 2005. Atmospheric aerosols: Composition, transformation, climate
  and health effects. Angew. Chem. Int. Ed. Engl. 44, 752--754.

\bibitem[{Prenni et~al.(2007)Prenni, Petters, Kreidenweis, DeMott, and
  Ziemann}]{Prenni2007}
Prenni, A., Petters, M., Kreidenweis, S., DeMott, P., Ziemann, P., 2007. {Cloud
  droplet activation of secondary organic aerosol}. J. Geophys. Res 112,
  D10223.

\bibitem[{Press et~al.(2007)Press, Teukolsky, Vetterling, and
  Flannery}]{NR2007}
Press, W.~H., Teukolsky, S.~A., Vetterling, W.~T., Flannery, B.~P., 2007.
  Numerical Recipes: The Art of Scientific Computing, 3rd Edition. Cambridge
  University Press.

\bibitem[{Riemer et~al.(2004)Riemer, Vogel, and Vogel}]{Riemer2004}
Riemer, N., Vogel, H., Vogel, B., 2004. Soot aging time scales in polluted
  regions during day and night. Atmospheric Chemistry and Physics 4,
  1885--1893.

\bibitem[{Riemer et~al.(2003)Riemer, Vogel, Vogel, and Fiedler}]{Riemer2003}
Riemer, N., Vogel, H., Vogel, B., Fiedler, F., 2003. Modeling aerosols on the
  mesoscale $\gamma$, {p}art {I}: {T}reatment of soot aerosol and its radiative
  effects. J. Geophys. Res. 108, 4601.

\bibitem[{Riemer et~al.(2009)Riemer, West, Zaveri, and Easter}]{Riemer2009}
Riemer, N., West, M., Zaveri, R., Easter, R., 2009. Simulating the evolution of
  soot mixing state with a particle-resolved aerosol model. J. Geophys. Res.In
  press.

\bibitem[{Rudich et~al.(2007)Rudich, Donahue, and Mentel}]{Rudich2007}
Rudich, Y., Donahue, N.~M., Mentel, T.~F., 2007. Aging of organic aerosol:
  Bridging the gap between laboratory and field studies. Annual Rev. Phys.
  Chem. 58, 321--352.

\bibitem[{Schell et~al.(2001)Schell, Ackermann, Binkowski, and
  Ebel}]{Schell2001}
Schell, B., Ackermann, I.~J., Binkowski, F.~S., Ebel, A., 2001. Modeling the
  formation of secondary organic aerosol within a comprehensive air quality
  model system. J. Geophys. Res. 106, 28275--28293.

\bibitem[{Svenningsson et~al.(2006)Svenningsson, Rissler, Swietlicki, Mircea,
  Bilde, Facchini, Decesari, Fuzzi, Zhou, M{\o}nster,
  et~al.}]{Svenningsson2006}
Svenningsson, B., Rissler, J., Swietlicki, E., Mircea, M., Bilde, M., Facchini,
  M., Decesari, S., Fuzzi, S., Zhou, J., M{\o}nster, J., et~al., 2006.
  {Hygroscopic growth and critical supersaturations for mixed aerosol particles
  of inorganic and organic compounds of atmospheric relevance}. Atmos. Chem.
  Phys 6, 1937--1952.

\bibitem[{Toner et~al.(2006)Toner, Sodeman, and Prather}]{Toner2006}
Toner, S., Sodeman, S., Prather, K., 2006. Single particle characterization of
  ultrafine and accumulation mode particles from heavy duty diesel vehicles
  using aerosol time-of-flight mass spectrometry. Environ. Sci. Technol. 40,
  3912--3921.

\bibitem[{Warner(1968)}]{Warner1968}
Warner, J., 1968. The supersaturation in natural clouds. J. Appl. Meteorol. 7,
  233--237.

\bibitem[{Weingartner et~al.(1997)Weingartner, Burtscher, and
  Baltensperger}]{Weingartner1997}
Weingartner, E., Burtscher, H., Baltensperger, H., 1997. Hygroscopic properties
  of carbon and diesel soot particles. Atmos. Environ. 31, 2311--2327.

\bibitem[{Zaveri et~al.(2008)Zaveri, Easter, Fast, and Peters}]{Zaveri2008}
Zaveri, R.~A., Easter, R.~C., Fast, J.~D., Peters, L.~K., 2008. {Model for
  Simulating Aerosol Interactions and Chemistry ({MOSAIC})}. J. Geophys. Res.
  113, D13204.

\bibitem[{Zaveri et~al.(2005{\natexlab{a}})Zaveri, Easter, and
  Peters}]{Zaveri2005b}
Zaveri, R.~A., Easter, R.~C., Peters, L.~K., 2005{\natexlab{a}}. A
  computationally efficient {M}ulticompoennt {E}quilibrium {S}olver for
  {A}erosols ({MESA}). J. Geophys. Res. 110, D24203.

\bibitem[{Zaveri et~al.(2005{\natexlab{b}})Zaveri, Easter, and
  Wexler}]{Zaveri2005a}
Zaveri, R.~A., Easter, R.~C., Wexler, A.~S., 2005{\natexlab{b}}. A new method
  for multicomponent activity coefficients of electrolytes in aqueous
  atmospheric aerosols. J. Geophys. Res. 110, D02210, doi:10.1029/2004JD004681.

\bibitem[{Zaveri and Peters(1999)}]{Zaveri1999}
Zaveri, R.~A., Peters, L.~K., 1999. A new lumped structure photochemical
  mechanism for large-scale applications. J. Geophys. Res. 104, 30387--30415.

\end{thebibliography}

\newpage

\begin{table}
  \begin{center}
    \begin{tabular}{c c c c c}
      \hline
      Initial/Background & $N$ ($\rm m^{-3}$) & $D_{\rm gn}$ ($\rm \mu m$)
      & $\sigma_{\rm g}$ (1)
      & Composition by mass \\
      \hline
      Aitken Mode & $3.2\cdot 10^{9}$  & 0.02                    & 1.45
      & 50\% $\rm (NH_4)_2SO_4$, 50\% POA \\
      Accumulation Mode & $2.9\cdot 10^{9}$  & 0.116                   & 1.65
      & 50\% $\rm (NH_4)_2SO_4$, 50\% POA\\
      \hline \\
      Emissions & $E$ ($\rm m^{-2}\, s^{-1}$)       & $D_{\rm gn}$ ($\rm \mu m$)
      & $\sigma_{\rm g}$ (1)
      & Composition by mass \\
      \hline
      Meat cooking & $9\cdot 10^{6}$      & 0.086                   & 1.9
      & 100\% POA \\
      Diesel vehicles & $1.6\cdot 10^{8}$ & 0.05                    & 1.7
      & 30\% POA, 70\% BC \\
      Gasoline vehicles& $5\cdot 10^{7}$  & 0.05                    & 1.7
      & 80\% POA, 20\% BC
    \end{tabular}
  \end{center}
  \caption{Initial and emitted aerosol distribution parameters. The
    initial aerosol distribution is also used as the background
    aerosol distribution. The percentages for the composition are by
    mass. $E$ is the area source strength of particle
    emissions. Dividing $E$ by the mixing height and
    multiplying by a normalized composition distribution gives the
    number distribution emission rate. \label{tab:aero_dat}}
\end{table}

\begin{table}
  \begin{center}
    \begin{tabular}{p{4.1cm} p{9cm}}
      \hline
      Terms & Description \\
      \hline
      \raggedright
      $N_{\rm f}(t)$,
      $N_{\rm a}(t)$
      &
      \raggedright
      Number concentration of fresh/aged BC-containing particles.
      \tabularnewline
      \raggedright
      $\dot N^{\rm emit}_{\rm f}(t)$,
      $\dot N^{\rm emit}_{\rm a}(t)$
      &
      \raggedright
      Gain rate of number concentration of fresh/aged BC-containing particles due to emission.
      \tabularnewline
      \raggedright
      $\dot N^{\rm dilution}_{\rm f}(t)$,
      $\dot N^{\rm dilution}_{\rm a}(t)$
      &
      \raggedright
      Loss rate of number concentration of fresh/aged BC-containing particles due to dilution.
      \tabularnewline
      \raggedright
      $\dot N^{\rm cond}_{\rm f \to a}(t)$,
      $\dot N^{\rm cond}_{\rm a \to f}(t)$
      &
      \raggedright
      Gain rate of number concentration of aged/fresh BC-containing particles due to 
      condensation or evaporation on fresh/aged particles.
      \tabularnewline
      \raggedright
      $\dot N^{\rm coag}_{\rm f}(t)$,
      $\dot N^{\rm coag}_{\rm a}(t)$
      &
      \raggedright
      Gain rate of number concentration of fresh/aged BC-containing particles from coagulation events.
      \tabularnewline
      \raggedright
      $\dot N^{\rm coag}_{\rm f \to f}(t)$,
      $\dot N^{\rm coag}_{\rm f \to a}(t)$
      &
      \raggedright
      Loss rate of number concentration of fresh BC-containing particles to coagulation 
      events resulting in fresh/aged particles.
      \tabularnewline
      \raggedright
      $\dot N^{\rm coag}_{\rm a \to a}(t)$,
      $\dot N^{\rm coag}_{\rm a \to f}(t)$
      &
      \raggedright
      Loss rate of number concentration of aged BC-containing particles to coagulation events 
      resulting in aged/fresh particles.
      \tabularnewline
      \raggedright
      $\dot N^{\rm density}_{\rm f}(t)$,
      $\dot N^{\rm density}_{\rm a}(t)$
      &
      \raggedright
      Gain rate of number concentration of fresh/aged BC-containing particles 
      due to air density changes.
      \tabularnewline
      \raggedright
      $\dot N^{\rm aging}(t)$,
      $\dot N^{\rm de-aging}(t)$
      &
      \raggedright
      Net transfer rate of fresh-to-aged/aged-to-fresh number concentration of BC-containing particles.
    \end{tabular}
  \end{center}
  \caption{\label{tab:term_def} Description of individual terms in
    equations~(\ref{eqn:num_dot_f}) and~(\ref{eqn:num_dot_a}). With the
    exception of $\dot N^{\rm density}_{\rm f}(t)$ and $\dot N^{\rm
      density}_{\rm a}(t)$ all of these terms must be non-negative. The
    same notation is for the terms in equations~(\ref{eqn:mass_dot_f})
    and~(\ref{eqn:mass_dot_a}) for mass concentration, and for the
    corresponding discrete
    equations~(\ref{eqn:n_dot_disc_f})--(\ref{eqn:m_dot_disc_a}). The
    discrete terms are expressed as a change in number or mass within a
    timestep, so that $\Delta n^{\rm cond}_{\rm f \to a}(t_{k-1},t_k)$
    is the number of BC-containing particles in the computational volume
    that change from fresh to aged due only to condensation during the
    timestep from time $t_{k-1}$ to $t_k$, for example. There are no discrete
    terms for air density changes as they are incorporated by changing
    the computational volume $V$.}
\end{table}

\begin{table}
  \begin{center}
    \begin{tabular}{p{1.6cm} p{1.6cm} p{1.6cm} p{4cm} p{2cm}}
      \hline
      Particle 1 & Particle 2 & Resulting particle & Non-zero loss terms                & Non-zero gain terms \\
      \hline
      fresh      & fresh      &  fresh             &  $\Delta n^{\rm coag}_{\rm f \to f}=2$ & $\Delta n^{\rm coag}_{\rm f}=1$ \\
      fresh      & fresh      &  aged              &  $\Delta n^{\rm coag}_{\rm f \to a}=2$ & $\Delta n^{\rm coag}_{\rm a}=1$ \\
      aged       & fresh      &  fresh             &  $\Delta n^{\rm coag}_{\rm a \to f}=1$, $\Delta n^{\rm coag}_{\rm f \to f}=1$ & $\Delta n^{\rm coag}_{\rm f}=1$ \\
      aged       & fresh      &  aged              &  $\Delta n^{\rm coag}_{\rm a \to a}=1$, $\Delta n^{\rm coag}_{\rm f \to a}=1$ & $\Delta n^{\rm coag}_{\rm a}=1$ \\
      aged       & aged       &  aged              &  $\Delta n^{\rm coag}_{\rm a \to a}=2$ & $\Delta n^{\rm coag}_{\rm a}=1$ \\
      fresh      & non-BC     &  fresh             &  $\Delta n^{\rm coag}_{\rm f \to f}=1$ & $\Delta n^{\rm coag}_{\rm f}=1$ \\
      fresh      & non-BC     &  aged              &  $\Delta n^{\rm coag}_{\rm f \to a}=1$ & $\Delta n^{\rm coag}_{\rm a}=1$ \\
      aged       & non-BC     &  fresh             &  $\Delta n^{\rm coag}_{\rm a \to f}=1$ & $\Delta n^{\rm coag}_{\rm f}=1$ \\
      aged       & non-BC     &  aged              &  $\Delta n^{\rm coag}_{\rm a \to a}=1$ & $\Delta n^{\rm coag}_{\rm a}=1$ \\
      \hline
    \end{tabular}
  \end{center}
  \caption{\label{tab:combinations} The different coagulation events and
    the resulting expressions for the loss and gain terms of the number
    of fresh and aged BC-containing particles. Similar expressions exist
    for mass changes $\Delta m$ in the particle-resolved model and for
    the number rates $\dot{N}$ and mass rates $\dot{M}$ in the
    continuous model.}
\end{table}
\clearpage
\newpage

\end{document}